\documentclass{jfm}
\usepackage{graphicx}
\usepackage{rotating,amsmath,float,upmath}
\usepackage{natbib}
\usepackage{color}
\usepackage{tikz}
\usepackage{array}
\newcolumntype{?}{!{\vrule width 1pt}}
\usepackage{booktabs} 
\usepackage{tabu}
\usepackage{bigints}

\ifCUPmtlplainloaded \else
  \checkfont{eurm10}
  \iffontfound
    \IfFileExists{upmath.sty}
      {\typeout{^^JFound AMS Euler Roman fonts on the system,
                   using the 'upmath' package.^^J}%
       \usepackage{upmath}}
      {\typeout{^^JFound AMS Euler Roman fonts on the system, but you
                   dont seem to have the}%
       \typeout{'upmath' package installed. JFM.cls can take advantage
                 of these fonts,^^Jif you use 'upmath' package.^^J}%
      }
  \else
  \fi
\fi

\ifCUPmtlplainloaded \else
  \checkfont{msam10}
  \iffontfound
    \IfFileExists{amssymb.sty}
      {\typeout{^^JFound AMS Symbol fonts on the system, using the
                'amssymb' package.^^J}%
       \usepackage{amssymb}%
       \let\le=\leqslant  \let\leq=\leqslant
       \let\ge=\geqslant  \let\geq=\geqslant
      }{}
  \fi
\fi

\ifCUPmtlplainloaded \else
  \IfFileExists{amsbsy.sty}
    {\typeout{^^JFound the 'amsbsy' package on the system, using it.^^J}%
     \usepackage{amsbsy}}
    {\providecommand\boldsymbol[1]{\mbox{\boldmath $##1$}}}
\fi

\def \bT {\mathbf{T}}

\def \bv {\mathbf{v}}

\def \t  {\theta}

{}

\def \bft{{\mathbf{f} \lp T \rp}}

\def \bt {\mathbf{t}}

\def \bn {\mathbf{n}}
\def \bu {\mathbf{u}}
\def \bv {\mathbf{v}}

\def \d    {\mathrm{d}}

\def \bT    {\mathbf{T}}

\def \d {\mathrm{d}}

\def \Uinf {U^{\infty}}

\def \buinf {\mathbf{U}^{\infty}}

\def \d {\mathrm{d}}

\def \bR {\mathbf{R}}
\def \bJ {\mathbf{J}}

\def \bI {\mathbf{I}}

\def \d {\mathrm{d}}

\def \be {\mathbf{e}}
\def \xin {\xi_{\perp}}
\def \xip {\xi_{\parallel}}
\def \bxc {\bx_{\mathrm{joint}}}
\def \nbxc {\bar{\bx}_{\mathrm{joint}}}

\def \bomgz {\bar{\omega}_z}

\def \tf {\theta_{f}}

\def \bh {\mathbf{h}}

\def\br{{\bar{\mathbf{r}}}}

\def\varepsilon{\epsilon}

\def \lp  {\left(}
\def \rp  {\right)}

\def \bx  {\mathbf{x}}

\def \bf {\mathbf{f}}
\def \bff {\bf^{(1)}}
\def \bfs {\bf^{(2)}}
\def \bft {\bf^{(3)}}

\def \bt {\mathbf{t}}
\def \btf {\bt^{(1)}}
\def \bts {\bt^{(2)}}

\def \br {\mathbf{r}}
\def \brf {\br^{(1)}}
\def \brs {\br^{(2)}}
\def \brt {\br^{(3)}}

\def \bJf {\bJ^{(1)}}
\def \bJs {\bJ^{(2)}}
\def \bJt {\bJ^{(3)}}

\def \bJf {\bJ^{(1)}}
\def \bJs {\bJ^{(2)}}
\def \bJt {\bJ^{(3)}}

\def \Uinf {U^{\infty}}

\def \hbar {\bar{h}}

\def \bI {\mathbf{I}}

\def\br{{\mathbf{r}}}

\title{Rotation of a low-Reynolds-number watermill: theory and simulations}

\def \bf {\mathbf{f}}

\author[L. Zhu and H. A. Stone]%
{Lailai Zhu$^{1,2}$ and Howard A. Stone$^{1}$%
  \thanks{Email address for correspondence: hastone@princeton.edu}}

\affiliation{$^1$Department of Mechanical and Aerospace Engineering, Princeton 
University, Princeton, New Jersey 08544, USA\\$^2$Linn\'{e} Flow Centre and 
Swedish e-Science Research Centre 
(SeRC), KTH 
Mechanics, Stockholm, SE-10044, Sweden}

\begin{document}

\maketitle 

\begin{abstract}
Recent experiments have demonstrated that small-scale rotary 
devices installed in a microfluidic channel can be  driven passively  by
the underlying flow alone without resorting to conventionally applied magnetic 
or electric fields. In this work, we conduct a theoretical and numerical study 
on 
such a flow-driven ``watermill'' at low Reynolds number, focusing on its 
hydrodynamic features. We model the watermill by a collection of equally-spaced
rigid rods. Based on  the 
classical resistive force (RF) theory and direct numerical simulations, we 
compute the watermill's instantaneous rotational velocity as a function of its 
rod number $N$, position and orientation. When $N \geq 4$, the RF theory 
predicts that the watermill's 
rotational velocity is
independent of $N$ and its orientation, implying the full rotational symmetry 
(of infinity order),
even though the geometrical configuration exhibits a lower-fold
rotational symmetry; the numerical solutions including hydrodynamic 
interactions show a weak dependence on $N$ and the orientation. In 
addition,
we adopt a dynamical system 
approach  to identify 
the equilibrium positions of the watermill and analyse their stability.
We further compare the  theoretically and numerically derived 
rotational velocities, which  agree with each other in general, while 
considerable discrepancy arises in certain configurations owing to the 
hydrodynamic interactions neglected by the RF theory. We confirm this 
conclusion by 
employing the RF-based asymptotic framework incorporating hydrodynamic 
interactions for a simpler watermill consisting of 
two or three rods and we show that accounting for hydrodynamic interactions can 
significantly enhance the accuracy of the theoretical predictions.

\end{abstract}

\section{Introduction}\label{sec:introduction}
In microfluidic devices, manipulation of the flow and the suspended phases 
such as cells, droplets/bubbles, macromolecules (e.g. DNAs), etc. is commonly 
needed. The flow manipulation 
includes mixing, pumping, valving, sensing and related operations. In order to 
achieve these 
functions, different strategies have been developed. One of the most intuitive 
approaches is to introduce into the microfluidic device a rotary element, 
whose rotation is achieved by applying an external 
electric~\citep{bart1992electric} or magnetic 
field~\citep{ahn1995fluid, 
dopper1997micro,ryu2004micro,agarwal2005programmable, van2015bidirectional}. 
On the other hand, such elements are also able to rotate passively without 
resorting to any external fields, but propelled by the underlying 
flow alone if they are placed asymmetrically with respect to the 
flow. This approach has been demonstrated by 
~\citet{zaki1994numerical} and ~\citet{day2000lubrication} where the latter was 
inspired by the experimental work of rotating an asymmetrically placed 
cylinder to
pump fluid in a duct~\citep{sen1996novel}. More recently,
~\citet{moon2015rotary} has succeeded to drive rotary 
microgears  by the underlying flow in a microfluidic channel. Each of 
their microgears consisted of eight paddles equally spaced in angle, and was 
fabricated and installed via in situ polymerisation based on flow lithography. 
The authors also showed that a pair of microgears was able to transmit the 
hydrodynamic torque from one gear to the other. Likewise, a similar flow-driven 
wheel
was implemented by ~\citet{attia2008modifications} (PhD thesis in French) 
as a flow sensor to measure the flow speed
in a microfluidic channel.

Motivated by such
microfluidic experiments, we hereby carry out a theoretical and numerical study 
on the low-Reynolds-number hydrodynamics of such flow-driven rotary 
devices 
resembling micro-scale ``watermills''.
We aim to provide design principles for their applications in microsystems.
After presenting the problem setup in Sec.~\ref{sec:problem}, we describe in 
Sec.~\ref{sec:meth}
the methodologies  including the classical resistive 
force (RF) theory and numerical methods. The results obtained are 
 compared in Sec.~\ref{sec:nohi}, which identifies the 
role of hydrodynamic interactions absent from the classical RF theory. 
Therefore,  
an improved RF theory taking hydrodynamic interactions into account is 
developed 
in Sec.~\ref{sec:hi_theo}  based on the recent 
theoretical work of ~\citet{man2016hydrodynamic}. We use the 
improved theory to solve the resistance and mobility problems of a 
simplified version of the watermill, and compare the theoretical predictions 
with the numerical results. Finally, we conclude and discuss our results in 
Sec.~\ref{sec:conclusion}.

\section{Problem setup}\label{sec:problem}
\begin{figure}
\centering
	
\includegraphics[scale=0.6]{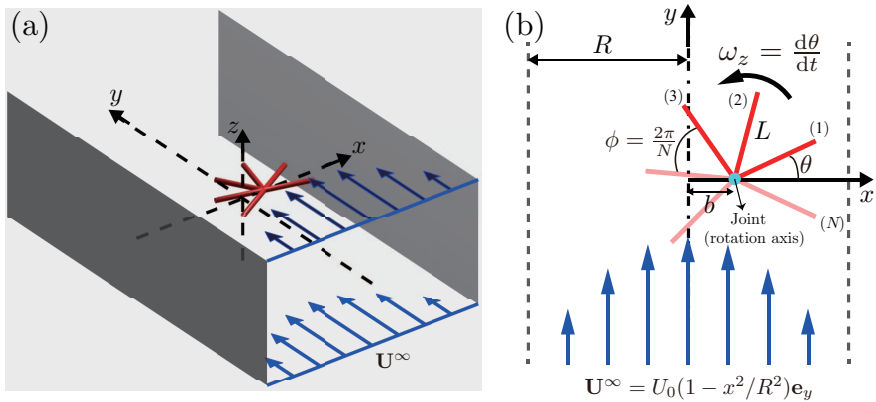}
		\caption{(Colour online) A watermill free to rotate in an 
underlying 
Poiseuille flow. (a) Sketch of an anchored, freely rotating 
watermill comprised of $N$ rigid cylindrical rods equally spaced in angle. 
The motion is driven 
by an unbounded Poiseuille flow $\buinf$. (b) The rotation axis of the 
watermill 
is aligned in
the $z$ direction and is away from the centre of the flow by a
distance $b$ in the $x$ direction.} 
\label{fig:sketch}
\end{figure}

We consider a watermill-like rotary device consisting 
of $N \geq 1$ cylindrical rods equally distributed in a plane in the 
azimuthal direction (see figure~\ref{fig:sketch}).
The angle between two successive rods is $\phi=2\pi/N$. All of the rods 
are jointed on a common end on the rotation axis of the device that is along 
the $z$ direction. Hence the device rotates 
in the horizontal $xy$ plane. The length of a rod is $L$, with its circular 
cross section of radius $a$, and the slenderness of the rod is defined as 
$\epsilon = a/L \ll 1$. 
We place the watermill in an unbounded Poiseuille flow $\buinf= 
U_0(1-x^2/R^2)\be_{y}$,
with the position $\bxc$ of the watermill's joint (rotation axis) away 
from the centre ($x=0$) of 
the flow 
domain by distance $b \geq 0 $, i.e., $\bxc = b \be_{x} $. Two 
nondimensional parameters
\begin{subequations}\label{eq:nondim}
\begin{align}
\beta & = b/L, \\
\rho & = R/L,
\end{align} 
\end{subequations}
are introduced to indicate the off-centre displacement of the watermill and
 the characteristic width of the channel flow, respectively.
The orientation of the watermill is indicated by the angle $\theta$ between 
the $1$-st rod 
(arbitrarily labelled without losing generality) and the 
$x$-axis. The rotational velocity $\omega_z = \d \theta/\d t$ of 
the watermill depends on its orientation $\theta \in [0, \phi]$. 

The dynamic viscosity of the fluid is $\mu$. We choose $U_0$, $L$, $L/U_0$, $\mu 
U_0/ L$, 
$\mu U_0 L$ 
and $\mu U_0 L^2$ as, respectively the characteristic velocity, length, time, 
stress,
force and 
torque.
Nondimensional quantities are denoted by $\;\bar{}\;$ from hereafter. We fix 
$\epsilon=0.02$ in this study.

\section{Methodologies}\label{sec:meth}
We carry out our study in the low-Reynolds-number flow regime and thus solve 
the 
Stokes equations. By employing the classical resistive force (RF) 
theory, we calculate the 
rotational velocities of a freely rotating watermill consisting of 
equally spaced rods. In Sec.~\ref{sec:nohi}, the results are
compared with those computed by direct numerical simulations of the Stokes 
equations. The 
comparisons indicate that inter-rod hydrodynamic interactions 
neglected by the RF theory play an important role in certain configurations. 
This feature thus motivates us to conduct a  theoretical study 
adopting the recently developed RF-based mathematical 
framework of ~\cite{man2016hydrodynamic} that accounts for hydrodynamic 
interactions; this ``RF-HI'' theory  will be
described in Sec.~\ref{sec:withhi}. The classical RF theory and numerical 
methods 
are documented in Sec.~\ref{sec:rf} and ~\ref{sec:nm}, respectively.

\subsection{Classical resistive force theory}\label{sec:rf}
We define an arclength $s \in [0, L]$
on each rod, with $s=0$ and $s=L$ corresponding to the joint and free end, 
respectively.
The RF theory 
dictates that the hydrodynamic force per unit length $\bf^{(k)}$ exerted by the 
fluid on the $k$-th rod is a function of the 
arclength $s$ according to
\begin{align}
 \bf^{(k)} (s) = -\bJ^{(k)}  \cdot \left[  \frac{\partial \br^{(k)}}{\partial 
t} 
- \buinf \lp 
\br^{(k)} \rp \right],
\end{align}
where  $\br (s)$ and $\bt(s) = \partial \br/\partial s$ 
denote, respectively, the local
coordinates and tangent of the rod's centreline; 
$\bJ (\bt) = \xin \bI + 
\lp \xip - \xin \rp \bt\bt$, 
where $\xin$ and 
$\xip$ are the drag coefficients for the motion of rod in the directions 
perpendicular 
and parallel to $\bt$. Note that $\xin = 2 \xip \approx 4\pi 
\mu 
/\ln{\epsilon^{-1}} $~\citep{lighthill1975mathematical}. 

The hydrodynamic torque exerted on the $k$-th rod about the joint is $\bT^{(k)} 
= \int_{0}^{L} \lp \br^{(k)} - \bxc \rp \times \bf^{(k)} \d s$. Since the 
watermill only rotates in the $xy$ plane, we only consider the $z$-component of 
the torque on the $k$-th rod, 
whose nondimensional value is 
\begin{align}\label{eq:non_tq1}
\bar{T}_z^{(k)} = \frac{1}{3 \rho^2 \ln{{\epsilon}}^{-1}} \left[ 6\lp 
\beta^2-\rho^2  
\rp \cos{\theta^{(k)}} + 8 \beta \cos^2{\theta^{(k)}} 
+ 3 
\cos^3{\theta^{(k)}} + 4\rho^2 \bar{\omega}_z \right],
\end{align}
where $\bar{\omega}_z$ is the nondimensional rotational velocity of the 
watermill; $\theta^{(k)}$ denotes the angle between the $k$-th rod and the 
$x$-axis and $\theta^{(1)}=\theta$. The total torque on the rotary 
device,  $ \bar{T}_{z} = 
\sum_{k=1}^{N} \bar{T}_z^{(k)}$, from which we find

\begin{align}
\displaystyle
 3  \rho^2 \lp \ln{{\epsilon}^{-1}} \rp \bar{T}_{z} = \sum_{k=1}^{N} \Bigg\{ 
\left[ 6 
\lp 
\beta^2-\rho^2 \rp 
+ \frac{9}{4}\right]\cos{\theta^{(k)}} + \\ \nonumber
4\beta\cos{2\theta^{(k)}} + 
\frac{3}{4}\cos{3\theta^{(k)}} \Bigg\} + 4N\lp \beta + \rho^2 
\bar{\omega}_z \rp.
\end{align}

Since the rods are equally spaced on a circle, we use the properties of 
roots of unity (detailed in appendix~\ref{sec:unit_root}) to obtain that
\begin{align}\label{eq:unitroot1}
   \sum_{k=1}^{N}\cos{\theta^{(k)}}  = 
\begin{cases}
    \cos{\theta},&  N=1\\
    0,              & N \ge 2,
\end{cases} 
\end{align}

\begin{align}\label{eq:unitroot2}
   \sum_{k=1}^{N}\cos{2\theta^{(k)}}  = 
\begin{cases}
    \cos{2\theta},&  N=1\\
    2\cos{2\theta},              & N = 2 \\
    0,              & N \ge 3,
\end{cases} 
\end{align}
and 
\begin{align}\label{eq:unitroot}
   \sum_{k=1}^{N}\cos{3\theta^{(k)}}  = 
\begin{cases}
    \cos{3\theta},&  N=1\\
    0,              & N = 2 \\
    3\cos{3\theta},              & N = 3 \\
    0,              & N \ge 4, 
\end{cases} 
\end{align}
so that the total torque $\bar{T}_z$ can be written as
\begin{align}
    3 \rho^2 \lp \ln{{\epsilon}^{-1}}\rp \bar{T}_{z}  = 
\begin{cases}
    [6(\beta^2-\rho^2)+\frac{9}{4}]\cos{\theta} + 
4\beta \cos{2\theta}+\frac{3}{4}\cos{3\theta} + 4(\beta+\rho^2\bar{\omega}_z),& 
 N=1\\
    8\beta \cos{2\theta}+8(\beta+\rho^2\bar{\omega}_z),              & N = 2 \\
    \frac{9}{4}\cos{3\theta}+12(\beta+\rho^2\bar{\omega}_z),              & N = 
3 \\
    4N(\beta+\rho^2\bar{\omega}_z),              & N \ge 4. 
\end{cases}
\label{eq:torquez}
\end{align} 

The rotational velocity of the freely rotating watermill can be obtained by
applying the torque-free condition $\bar{T}_z=0$ and we find

\begin{align}
   \bar{\omega}_z=  
\begin{cases}
    -\cos\theta[6(\beta^2-\rho^2)  + 8\beta\cos{\theta}+ 
3\cos^2{\theta}]/\lp 4\rho^2\rp,&  N=1\\
    -2\beta\cos^2{\theta}/\rho^2,              & N = 2 \\
    -\left[16\beta+3\cos{3\theta}\right]/\lp 16\rho^2\rp,              & N = 3 
\\
    -\beta/\rho^2,              & N \ge 4. 
\end{cases}
\label{eq:omegaz}
\end{align}

\subsection{Numerical methods}\label{sec:nm}
To determine the rotational velocity of the watermill numerically, we carry
out three-dimensional direct numerical simulations (DNS) 
based on a commercial finite-element method (FEM) solver COMSOL. We have 
experience in performing COMSOL simulations for viscoelastic flows, 
e.g.~\cite{pak2012micropropulsion} where the propulsion of two touching 
rotating spheres
in viscoelastic fluids was investigated and
the numerical results were in excellent agreement with the asymptotic analysis 
in the small Deborah number regime. It is worth noting that prior studies 
of interacting slender bodies in viscous flows have 
adopted other numerical 
implementations~\citep{yamamoto1995dynamic,  
ross1997dynamic, saintillan2007orientational, nazockdast2017fast}.

Since we assume that the Reynolds number $Re=\rho_{f} \Uinf L/\mu$ is 
small ($\rho_f$ denotes the fluid density), inertia effects are negligible, 
and we solve 
the nondimensional steady Stokes equations 
\begin{align}
 -\nabla \bar{p} + \nabla^2 \bar{\bu} & = \mathbf{0},\\
 \nabla \cdot \bar{\bu} & =0.
\end{align}
For a channel flow, we impose a Dirichlet boundary condition (BC) with the 
Poiseuille flow profile in the inlet of the domain and a constant pressure BC at 
the outlet. Utilising the mirror symmetry of the setup, we only need to consider 
the upper half ($z \geq 0 $) of the domain
by applying a symmetry BC at the $z=0$ plane. We adopt the same symmetry
BC at the $z=H_z/2$ plane, which effectively corresponds to an 
array of 
watermills equally spaced along the $z$ 
direction by distance $H_z$.
No-slip BCs are specified on the two lateral walls at $x =  \pm R$. 
Since both the RF and RF-HI theories are derived for unbounded flows, i.e. 
without accounting for wall effects, the 
boundedness of the computational domain needs to be considered carefully for 
a 
reasonable comparison of the theoretical and numerical results. We choose the 
length $H_x$ of the domain equal to $100L$. To mitigate the confinement 
effects of the lateral walls,
the characteristic width  of the channel flow $\rho=5$ is used in most of our 
cases. We have varied the distance $H_z$ in the range $\in [0.25, 4]L$
and find $H_z=2L$ is 
sufficiently large to guarantee that the hydrodynamic interaction between
the watermill and its mirror image about the symmetry BC at $z=H_z/2$ is 
negligible.

The BC imposed at position $\bx_{w}$ on the surface $S_w$ of 
the watermill is 
\begin{align}
 \bar{\bu} \lp \bar{\bx}_{w}  \rp = \bar{\omega}_z \be_z \times 
(\bar{\bx}_{w} - \nbxc ),
\end{align}
where $\nbxc=\beta \be_x$ denotes the nondimensional position of the 
joint of the watermill. Note that the rotational velocity 
$\bar{\omega}_z$ 
is  an unknown and it is solved together with the flow field $\lp 
\bar{p}, \bar{\bu}\rp$
by incorporating the  constraint of zero hydrodynamic torque on the watermill
\begin{align}
 \int_{S_w} (\bar{\bx}_{w} - \nbxc ) \times \left[ \lp -\bar{p}\bI + \nabla 
\bar{\bu} + \nabla \bar{\bu}^{T}    \rp \cdot \bn_w \right] \mathrm{d}S= 
\mathbf{0}, 
\end{align}
where $\bn_w$ denotes the unit normal vector on $S_w$.

The numerical setup is validated for a resistance and a mobility problem. For 
the resistance problem, we consider a cylindrical rod rotating at a 
constant velocity $\omega \be_z$ about one of its ends, with its revolution 
axis 
on the $z=0$ plane. The hydrodynamic torque $T_z/\lp\mu \omega L^3 \rp$ 
calculated numerically agrees well with the RF predictions for varying 
slenderness $\epsilon$, as shown in table~\ref{tab:toq_rod_shear}.
For the mobility problem, we consider the free rotation of a spheroid in a 
shear flow, $\mathbf{U}^{\infty}=-\dot{\gamma} y \be_x$, where the revolution 
axis of the spheroid is on the shear  plane ($z=0$).  We define $2\kappa$ as
the length of the axis of revolution scaled by the radius of a sphere with 
the same volume, and $\alpha$ as the angle between the revolution axis and
the $x$-axis. We compute the instantaneous rotational velocity $\Omega = 
\Omega_z \be_z $ of the spheroid as a function of $\kappa$ and $\alpha$, 
and validate the results against the analytical 
theory~\citep{jeffery1922motion}, $\Omega_z /\dot{\gamma}= 0.5\left[ 1- 
(\kappa^3-1)/(\kappa^3+1)\cos{2\alpha} \right]$. The comparison shown in 
table~\ref{tab:omg_ellip} shows a maximum discrepancy of below $1\%$.

{
\tabulinesep=1.5mm
\begin{table}
\centering
\begin{tabu}
spread 0pt {|X[3l]?X[c]|X[c]|X[c]|X[c]|X[c]|} 
\tabucline[\heavyrulewidth]-
             &$\epsilon=0.01$                  
&$\epsilon=0.02$ 
&$\epsilon=0.03$ &$\epsilon=0.04$ &$\epsilon=0.05$ \\ \tabucline-
$T_z/\mu \omega L^3$ (RF theory)       & 0.9096 & 1.0707 & 1.1946 & 1.3013 & 
1.3983 \\
$T_z/\mu \omega L^3$ (Simulation)    & 0.9093 & 1.0758 & 1.1998 & 1.3088 & 
1.3982 \\  
\tabucline[\heavyrulewidth]-
\end{tabu}
\caption{The nondimensional hydrodynamic torque $\bar{T}_z = T_z/\lp \mu \omega 
L^3 \rp$ exerted on a single rod of varying slenderness $\epsilon$ which 
rotates about one of its ends at a prescribed rotational velocity $\omega$
 in the $xy$ plane. }
\label{tab:toq_rod_shear}
\end{table}
}

{
\tabulinesep=1.5mm
\begin{table}
\centering
\begin{tabu}
spread 0pt {|X[2l]?X[c]X[c]X[c]X[c]|} 
\tabucline[\heavyrulewidth]-
   $\kappa=2$ (Prolate)            &$\alpha=0^{\circ}$                  
&$\alpha=30^{\circ}$ 
&$\alpha=60^{\circ}$ &$\alpha=90^{\circ}$  \\ \tabucline-
$\Omega_z/\dot{\gamma}$ (Theory)       & 0.1111 & 0.3056 & 0.6944 & 
0.8889  \\
$\Omega_z/\dot{\gamma}$ (Simulation)    & 0.1055 & 0.3025 & 0.6974 & 0.8945  \\ 
\tabucline[\heavyrulewidth]-
\tabucline[\heavyrulewidth]-
   $\kappa=0.6$ (Oblate)            &$\alpha=0^{\circ}$                  
&$\alpha=30^{\circ}$ 
&$\alpha=60^{\circ}$ &$\alpha=90^{\circ}$  \\ \tabucline-
$\Omega_z/\dot{\gamma}$ (Theory)       & 0.8224 & 0.6612 & 0.3388 & 
0.1776  \\
$\Omega_z/\dot{\gamma}$ (Simulation)    & 0.8293 & 0.6646 & 0.3354 & 0.1705  \\ 
\tabucline[\heavyrulewidth]-
\end{tabu}
\caption{The nondimensional rotational velocity $\Omega_z/\dot{\gamma}$ of a 
prolate ($\kappa=2$) and an oblate ($\kappa=0.6$) spheroid in shear flow 
$\mathbf{U}^{\infty}=-\dot{\gamma} y \be_x$ as a function
of the angle $\alpha$ between its axis of revolution and the 
streamwise 
($x$) direction. 
The theoretical results are from \citet{jeffery1922motion}. }
\label{tab:omg_ellip}
\end{table}
}

\section{Results: RF without hydrodynamic interactions}\label{sec:nohi}

\subsection{One individual rod: $N=1$}
\begin{figure}
\centering
	\includegraphics[scale=0.5]{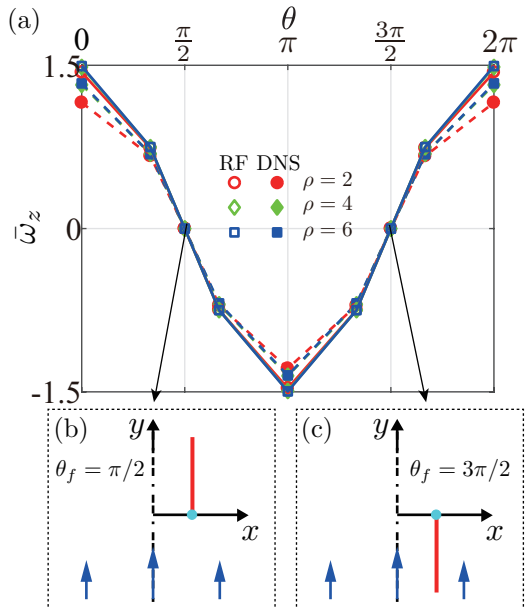}
		\caption{(Colour online) (a) Instantaneous rotational velocity 
$\bomgz$ of a single rod 
with an 
off-centre displacement $\beta=1/8$ in a channel flow. Empty and solid symbols 
denote, respectively the results of resistive theory (RF) and DNS for three 
characteristic widths of flow
$\rho=2$ (circle), $\rho=4$ (diamond) and $\rho=6$ (square). The 
configurations of the rod corresponding to the fixed points $\tf = \pi/2$ and 
$\tf = 
3\pi/2$ are depicted in (b) and (c), respectively. }  
\label{fig:1rods_rf}
\end{figure}

We first investigate one  rod and  plot its instantaneous rotational velocity 
$\bomgz$ 
as a 
function of its orientation $\theta$ in figure ~\ref{fig:1rods_rf}(a). The
off-centre displacement $\beta=1/8$ is fixed, and three characteristic channel 
widths
$\rho=2, 4$ and $6$ are studied.  In the most 
confined case $\rho=2$, the RF results
deviate with the DNS data when $\theta=0$, where the rod is 
oriented towards 
the nearest lateral wall; the relative difference between the two is  
approximately $24\%$. 
This result
is expected considering 
the rather strong confinement. When $\rho=4$ and $6$, the maximum 
discrepancy between the RF and DNS results is below $10\%$. 
We thus conclude that 
the effect of this level of confinement can be neglected and will not be 
considered further. 

We can consider $\d{\theta}/\d t = \bomgz\lp \t \rp$ as a one-dimensional
dynamical system and investigate the fixed points $\tf$ satisfying 
$\bomgz(\tf)=0$ and their stability.
Noticing that $\beta+1 \le \rho$ because the rod cannot penetrate the side 
walls 
and $\cos \t  \in [-1, 1]$, and using equation~(\ref{eq:omegaz})
we observe that 
$[6(\beta^2-\rho^2)  + 
8\beta\cos{\t} + 3\cos^2{\t}] > 0$. 
Therefore, $\cos {\tf } = 0$ leads  to the two fixed points $\tf = \pi/2$, 
i.e., the rod is aligned with the flow direction (figure~\ref{fig:1rods_rf}(b)); 
and $\tf  =3\pi/2$, 
the rod is oriented opposite to the flow direction 
(figure~\ref{fig:1rods_rf}(c)). 
Their
stability are dictated by the sign of the slope $\d \bomgz/\d \t 
|_{\tf}$. We hence observe that the former/latter fixed point is 
stable/unstable because $\d \bomgz/\d \t |_{\tf=\pi/2} < 0$ and 
$\d \bomgz/\d \t |_{\tf=3\pi/2} > 0$ as also is indicated in
figure~\ref{fig:1rods_rf}(a). We  conclude that the watermill
consisting of one rod will adopt a steady equilibrium
position in the channel flow. 

\subsection{Two rods: $N=2$}
\begin{figure}
\centering
	\includegraphics[scale=0.5]{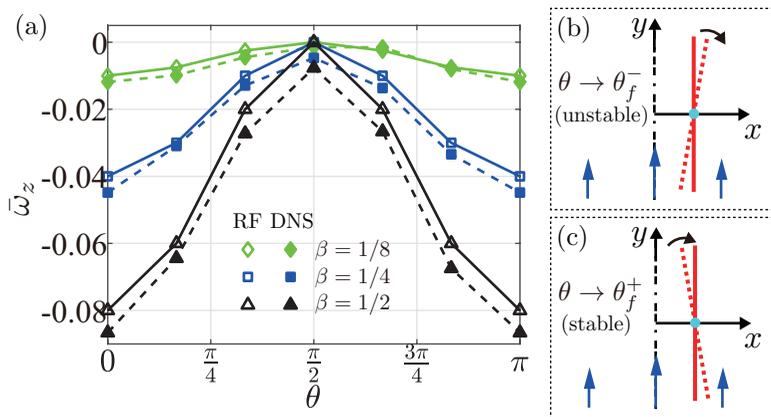}
		\caption{(a) Instantaneous rotational velocity 
$\bomgz$ of a $N=2$ 
watermill with an off-centre displacement $\beta=1/8$ (diamond), 
$\beta=1/4$ (square) and $\beta=1/2$ (triangle) in a channel with a
characteristic width $\rho=5$. (b) and (c) indicate, respectively, 
the 
unstable ($\theta \rightarrow {\tf}^{-}$) and stable ($\theta 
\rightarrow {\tf}^{+}$) configurations of the half-stable fixed 
point $\tf = \pi/2$. }  
\label{fig:2rods_rf}
\end{figure}

We now consider a watermill consisting of two rods separated by $\pi$,
which of course is equivalent to a single rod of length $2L$ with its centre
jointed at the rotation axis. The geometry has the rotational $2$-fold
symmetry
about the joint and hence its rotational 
velocity
preserves the periodicity $\bomgz(\t) = \bomgz(\t + \pi)$. 
Figure~\ref{fig:2rods_rf}(a) shows $\bomgz(\t)$ for $\t \in [0, \pi]$  of the 
watermill displaced by three off-centre distances $\beta$ for a fixed 
channel width $\rho = 5$. The rotational velocity reveals the mirror symmetry 
about $\theta = \pi/2$. This result is indicated by the analytical expression 
equation~(\ref{eq:omegaz}) when $N=2$, which also reflects the reversibility of 
the Stokes flow. The RF theory agrees reasonably well with the DNS data 
when the offset $\beta=1/4$ and $1/2$; the $\theta$-averaged relative 
difference  between the theory and DNS is around $14\%$. This difference however increases 
to $27\%$ for $\beta=1/8$, because the absolute value of $\bomgz$ becomes 
small and slight differences can introduce large relative deviations.
It is 
worth noting that the slight asymmetry of the DNS data about $\theta = \pi/2$  
reveals  the role of weak confinement captured by the simulations.

By setting $\bomgz(\tf) =0 $, we identify the only fixed point $\tf = 
\pi/2$ representing the configuration when two rods are perfectly aligned with 
the 
flow direction. Since the slope $\d \bomgz/\d \t |_{\tf} = 0$, 
this fixed point is neither stable or unstable, but is regarded as 
half-stable~\citep{strogatz2014nonlinear}, being unstable when 
$\theta \rightarrow  {\tf}^{-}$ (positive slope) and stable when 
$\theta \rightarrow  {\tf}^{+}$ (negative slope). The stability of
this equilibrium position thus depends on the sign of 
perturbations: subject to a negative perturbation $\delta \t<0$, the system 
would rotate half a circle before it recovers to the equilibrium state (see 
figure~\ref{fig:2rods_rf}(b)); otherwise, it is  
immediately stabilised to the equilibrium state  (see 
figure~\ref{fig:2rods_rf}(c)). In fact, this half-stability is in analogy with 
that 
of a prolate particle following Jeffery's orbit in 
the shear plane when  the particle's axis of revolution is 
along the flow direction~\citep{jeffery1922motion}.
The results imply that the two-rod 
watermill cannot be used in practice for flow manipulation on demand.

\subsection{Three rods: $N=3$}
\begin{figure}
\centering
	\includegraphics[scale=0.45]{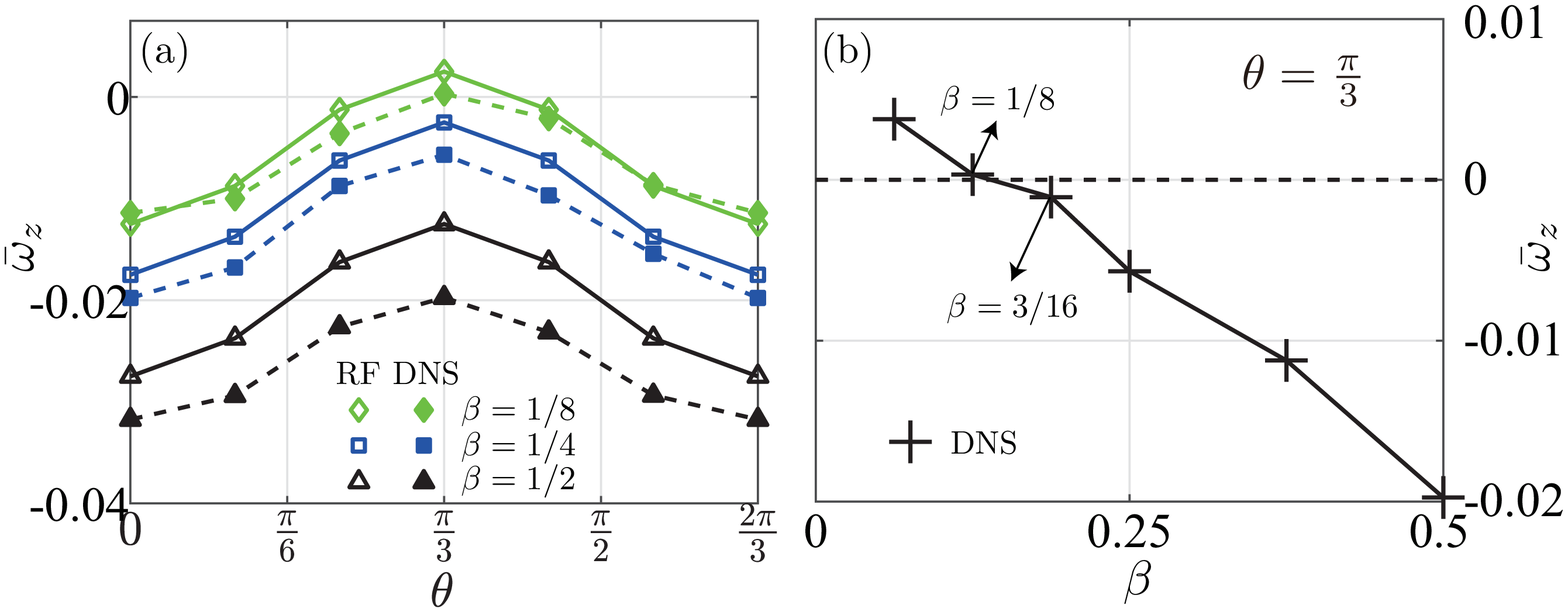}
		\caption{(Colour online)  (a) Instantaneous rotational velocity 
$\bomgz$ of a $N=3$ watermill 
with an off-centre displacement $\beta=1/8$ (diamond), 
$\beta=1/4$ (square) and $\beta=1/2$ (triangle) in a channel flow of a 
characteristic width $\rho=5$. (b) DNS results of  $\bomgz$  versus 
$\beta$ for a fixed orientation $\theta = \pi/3$.
}
\label{fig:3rods_rf}
\end{figure}

We present the RF-based rotational velocity $\bomgz 
= - \left[ 16\beta+3\cos(3\theta) \right] /16\rho^2$ (see 
equation~(\ref{eq:omegaz})) and
its DNS counterpart in figure~\ref{fig:3rods_rf}(a) for a three-rod
watermill. The solution is characterised 
by the $3$-fold
rotational symmetry about the joint, so the 
rotation 
of the watermill satisfies the periodic condition $\bomgz \lp \t \rp = \bomgz 
\lp \t + 2\pi/3 \rp $. The agreement between the RF and DNS results is in general less 
favourable compared to the cases of $N=1$ and $2$; the relative difference 
between them is about $23\%$, $19\%$ and $20\%$ for $\beta=1/8$, $1/4$ and 
$1/2$, respectively. When $\beta=1/2$, the 
magnitude of $\bomgz$ is systematically underestimated  by 
the RF theory. 
When $\beta = 1/4$ or $1/2$, $\bomgz$ is 
negative regardless
of the watermill orientation $\theta$, so the watermill rotates continuously. 
However, when $\beta=1/8$, $\bomgz$ is 
positive only as $\t$ is close to $\pi/3$. In fact, the analytical 
expression indeed indicates that the fixed-point solutions satisfying 
$\cos{(3\theta)} = -16\beta/3$ appear when $\beta $ is less than or equal to a 
critical value $\beta_{\ast} 
= 3/16 $; otherwise when $\beta > \beta_{\ast}$,  $\bomgz (\t)<0$  
for any $\t$. In the former case, we can identify only one fixed point $\tf = 
[2\pi - \arccos{(-16\beta/3)}]/3$ and it is stable.
The DNS results for $\bomgz$ reaches 
its maximum when $\theta=\pi/3$, which is consistent with the RF prediction. In 
figure~\ref{fig:3rods_rf}(b), we plot
 $\bomgz|_{\theta = \pi/3}$ given by DNS as a function of the offset $\beta$. 
The results 
show
that the critical offset is slightly above $1/8$, qualitatively confirming the 
theoretical prediction $\beta_{\ast} 
= 3/16 $. We conclude that in order to continuously rotate, a three-rod 
watermill needs to be placed a critical distance away from the flow centre.

\subsection{More than three rods: $N \ge 4$}

\begin{figure}
\centering
\hspace{-2.5em}
\includegraphics[scale=0.34]
{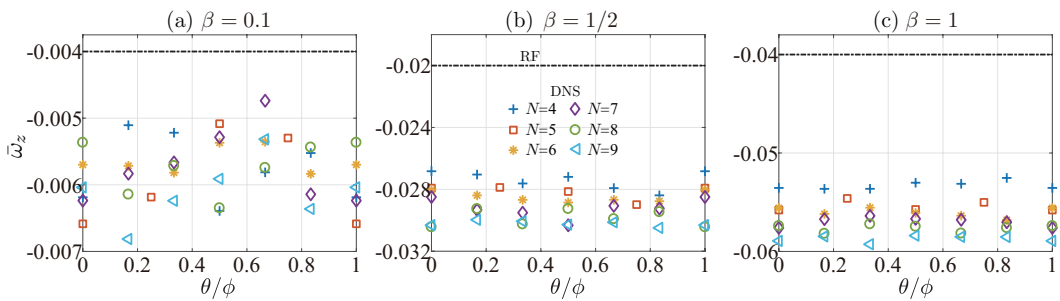}
		\caption{(Colour online)  Instantaneous rotational 
velocity 
$\bomgz$ of a watermill with an off-centre displacement 
(a) $\beta=0.1$, (b) $\beta=1/2$ and (c) $\beta=1$ in a channel with a 
characteristic width $\rho=5$. The horizontal dashed line corresponds to 
the RF prediction $\bomgz= -\beta/\rho^2$ and the symbols represent 
the orientation-discrete  DNS results for the watermill consisting of $N=4$ 
(cross), $5$ (square), $6$ (star), 
$7$ (diamond), 
$8$ (circle) and $9$ (triangle) rods.}  
\label{fig:multirods_rf}
\end{figure}

An important observation from the simple RF theory is that as long as $ N \geq 
4 
$, the rotational velocity of the watermill $\bomgz = -\beta / \rho^2$ is 
independent of $N$ or its orientation $\theta$  (see 
equation~(\ref{eq:omegaz})). This striking independence stems from the
 torque calculation that involves linear combinations of $
\cos\theta$, $\cos 2 \theta$ and $\cos 3\theta$ as indicated by
equation~(\ref{eq:torquez}), which is the mathematical reason that leads to the 
$N \geq 4$ threshold for a rotation speed independent of $N$ (see 
appendix~\ref{sec:unit_root} for the details). This prediction is 
qualitatively  confirmed
by the comparison between the RF and DNS data in figure~\ref{fig:multirods_rf}
for $N \in [4, 9]$ with three off-centre distances $\beta=0.1$, $1/2$ and $1$.
Due to the periodicity of $\bomgz$ in $\theta$, as demonstrated in the previous 
sections, the rotational velocity is plotted versus $\theta/ \phi $ to 
ease the comparison between different $N$. Figure~\ref{fig:multirods_rf}  shows 
that the  RF results agree with the
numerical data qualitatively. Compared to the independence of $\bomgz$ on 
$\theta$ predicted by RF, the DNS suggests that $\bomgz$ exhibits weak
dependence on the 
orientation $\theta$ at $\beta=0.1$ while its variation  in 
$\theta$ is negligible at $\beta=1/2$ and $1$. We also observe that the RF 
theory underestimates $\bomgz$ systematically compared to the DNS data, resulting in a 
relative difference of around $30\%$. Another 
important observation is that the magnitude of $\bomgz$ increases in general 
with the number $N$ of rods.

We infer that the underestimation of RF theory might be attributed to its 
lack of accounting for the inter-rod hydrodynamic interactions, which promotes 
the 
rotation of the watermill. Intuitively, the hydrodynamic interactions 
 depend significantly
on the separation distance (indicated by $\phi$) between every two 
successive rods and hence it should become stronger with decreasing $\phi$ 
(increasing $N$). This intuition might also explain the positive 
relation between $\bomgz$ and $N$ that is unaccounted for by the RF theory. 
We address these ideas further in the next section.

\section{RF considering hydrodynamic interactions}\label{sec:withhi}
\subsection{Theoretical framework}\label{sec:hi_theo}
\begin{figure}
\centering
	\includegraphics[scale=0.6]{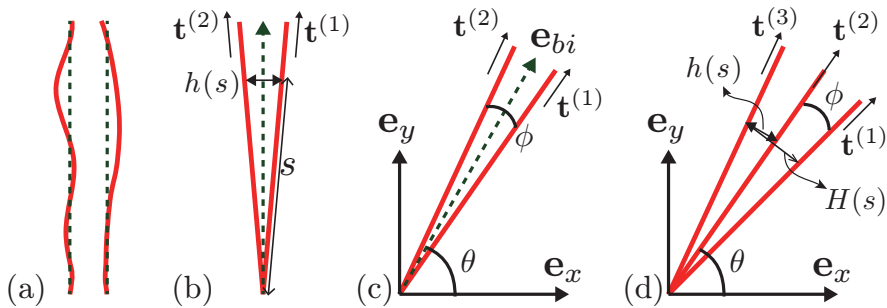}
		\caption{(Colour online) Sketch of two interacting 
rods/filaments. 
(a) The configuration of two filaments studied 
by
~\citet{man2016hydrodynamic}, where the dashed lines denote their initial, 
undeformed states. The 
rods are aligned (b) in the $\be_{bi} = \be_y$ direction  and (c) in an 
arbitrary direction 
$\be_{bi} = \cos{\theta} \be_x  + \sin{\theta} \be_y$; the dashed arrow
indicates the internal bisector of the two rods. (d): the configuration of 
three rods.}  
\label{fig:inter_2rod}
\end{figure}

The above results suggest that there are configurations where it is important 
to take hydrodynamic interactions into account in order to provide a more 
accurate prediction of the rotational velocities than the classical RF theory. 
As far as we know, theoretical efforts have been reported to address the 
hydrodynamic
interactions between a slender body and a 
wall~\citep{de1973low,russel1977rods,barta1988slender} or 
two walls~\citep{katz1974propulsion}. 
Based on 
the classical RF theory, we hereby make  an attempt by utilising 
a theoretical framework~\citet{man2016hydrodynamic} recently developed for 
two interacting weakly deformable filaments that are in parallel in their 
relaxed 
state (see figure~\ref{fig:inter_2rod}(a)). As a first step, we will consider a 
simplified configuration rather than the original setup, focusing 
on two or three rods, \textit{i.e.} $N\leq 3$; further, the separation angle 
$\phi$ is independent of $N$ and is tuned arbitrarily to be much smaller 
than $2\pi/N$, in contrast to the value $\phi=2\pi/N$ of the 
original setup. 

We start with a two-rod watermill and consider 
 the limit that $\epsilon \ll \phi \ll 1$ (see 
figure~\ref{fig:inter_2rod}(b) 
and (c)). 
This limit indicates 
first that the two rods are approximately parallel,  \textit{viz.} 
$\btf 
\approx \bts \approx \be_{bi}$, where $\be_{bi}$ is the orientation of 
their internal bisector; 
second, their typical separation distance $h \sim L \phi $ is 
much smaller than $L$ but much larger than the radius $a$, namely 
$a \ll h \ll L$. Note that in the near-hinge segment with sufficiently small 
arclength $s \leq a$, the latter approximation is violated. However, the 
introduced 
errors of
computing the corresponding hydrodynamic torque is rather limited, 
because of the very small prefactor, namely the distance $s$ away from the 
hinge.

The hydrodynamic force per 
unit length $\bf^{(k)}$ exerted by the 
fluid on the two rods is 
\begin{subequations}\label{eq:sbt}
\begin{align}
 \bff  -\bJ^{(1)} \cdot \bv^{(2)\rightarrow(1)} = -\bJ^{(1)} \cdot 
\left[ \frac{\partial \brf}{\partial t}-\buinf 
(\brf)\right], \\
 \bfs -\bJ^{(2)} \cdot \bv^{(1)\rightarrow(2)} = -\bJ^{(2)} \cdot 
\left[ \frac{\partial \brs}{\partial t}-\buinf 
(\brs)\right],
\end{align} 
\end{subequations}
where $\bv^{(2)\rightarrow(1)}$ denotes the velocity  induced 
by the $2$-nd rod  to the $1$-st rod. Following 
~\citet{man2016hydrodynamic}, we first 
assume that the orientation of the two rods  $\be_{bi}$ is parallel with 
$\be_y$ (see figure~\ref{fig:inter_2rod}(b)). 
The induced velocity 
$\bv^{(2)\rightarrow(1)}$   can be  integrated asymptotically
in the limit of $a \ll h \ll L$ ~\citep{man2016hydrodynamic},
\begin{align}\label{eq:v2to1_org}
 \bv^{(2)\rightarrow(1)} =\frac{1}{4\pi\mu}\ln{\lp\frac{h(s)}{L}\rp}(\bI + 
\be_y \be_y) \cdot \bfs(s), 
\end{align} 
where $\bh (s) = \brf (s) - \brs (s)$ denotes the local displacement vector 
between the two points $\brf(s)$ and $\brs(s)$ on the  rods' centrelines, and 
$h(s) = |\bh(s)| = 2 s \sin{(\phi/2)}$. 
We will retain the functional form with the sine, so as to possibly see 
whether this approach might work for large $\phi$ values, though  recognise 
that the self-consistency for small $\phi$ implies $\sin(\phi/2) \sim \phi/2$.

Because $\btf \approx \bts \approx \be_y$, we obtain $\bJ^{(1)} (\btf) \approx 
\bJ^{(2)} (\bts) \approx 
\bJ (\be_y) = \xi_{\perp} \lp  \bI - \be_y \be_y/2 \rp $. Thus, the force 
density
involving the induced velocity can be reformulated as 
\begin{align}\label{eq:J_dot_v}
 -\bJ^{(1)} \cdot \bv^{(2) \rightarrow (1)} =  \lambda \lp s \rp \bfs,
\end{align} 
where $\lambda \lp s \rp=\ln{\lp h(s)/L \rp}/\ln{\epsilon}$.
An important observation is that equation~(\ref{eq:J_dot_v})
does not depend on the orientation $\be_{bi}$ of the two rods. 
Because of
this independence from $\be_{bi}$, by considering a general orientation 
$\be_{bi} 
= \cos{\theta} \be_x  + \sin{\theta} \be_y$ (see 
figure~\ref{fig:inter_2rod}(c)), 
we can always rewrite equation~(\ref{eq:sbt}) and obtain a linear system with 
its left-hand side independent of $\t$ as
\begin{subequations}\label{eq:hi_2rod}
\begin{align}
  \bff + \lambda(s) \bfs   = -\bJf \cdot 
\left[ \frac{\partial \brf}{\partial t}-\buinf 
(\brf)\right], \\
  \bfs + \lambda(s) \bff   = -\bJs \cdot 
\left[ \frac{\partial \brs}{\partial t}-\buinf 
(\brs)\right].
\end{align} 
\end{subequations}

We note that equation~(\ref{eq:hi_2rod}) is slightly different from equation 
(25) of ~\citet{man2016hydrodynamic}. In their work, they aimed 
to model two flexible filaments initially undeformed and oriented parallel 
to
the $\be_y$ direction (see figure~\ref{fig:inter_2rod}(a)). Once deformed, they 
become slightly misaligned,  and 
$\bt^{(k)}(s)$ and $\bJ^{(k)} (\bt^{(k)}(s))$, with $k=1, 2$, on their 
centrelines, both vary with the arclength $s$.
Adopting the long-wavelength approximation, $\bJ^{(k)} 
 \approx \bJ^{(k)} \lp \be_y \rp$ is assumed; that said, 
they took the initial direction $\be_y$ of the filaments to 
approximately calculate $\bJ^{(k)}$, neglecting the influence of deformation.
Similar to 
~\citet{man2016hydrodynamic}, we 
compute $\bJ^{(k)}$ on the left-hand side of equation~(\ref{eq:sbt}) based 
on $\be_{bi}$, the approximate 
orientation of the two rods, hence we have $\bJ^{(k)} 
 \approx \bJ^{(k)} \lp \be_{bi} \rp$. This step facilitates the expression of  
$-\bJ^{(k)} \cdot \bv^{(j) \rightarrow (k)}$ as a function of local force 
density $\bf^{(j)}$, with $k \neq j$. Therefore, maintaining generality in 
geometry, the 
orientations of our rigid rods $\bt^{(k)}$ are well defined as
\begin{align}
 \btf = \cos{\lp \t - \phi/2 \rp} \be_x + \sin{\lp \t - \phi/2 \rp} \be_y, 
\nonumber \\
 \bts = \cos{\lp \t + \phi/2 \rp} \be_x + \sin{\lp \t + \phi/2 \rp} \be_y,
\end{align}
which are used to compute $ \bJ^{(k)}(\bt^{(k)})$ on the right-hand 
side of equation~(\ref{eq:sbt}). These steps lead to a discrepancy in computing 
$\bJ^{(k)}$ on the left- and right-hand sides of equation~(\ref{eq:sbt}), which 
vanishes  asymptotically when $\phi \rightarrow 0$.  In the small $\phi$ 
limit,
our derivations are consistent.

By inverting the matrix of equation~(\ref{eq:hi_2rod}), we obtain the force 
density $\bf^{(k)}$
\begin{align}
 \bf^{(k)} (s) = -\frac{1}{1 - \lambda^2(s)}  \lp \bJ^{(k)} \cdot \left[ 
\frac{\partial \br^{(k)}}{\partial t}  - \buinf 
(\br^{(k)}) \right] - \lambda(s) \bJ^{(j)}  \cdot \left[\frac{\partial 
\br^{(j)}}{\partial 
t}-\buinf 
(\br^{(j)})\right] \rp, k \neq j.
\end{align}

The above approach has been extended to the case of three rods (see 
figure~\ref{fig:inter_2rod}(d)). 
By considering
the mutual hydrodynamic interactions between every pair of two rods, we obtain 
the
linear system as
\begin{subequations}\label{eq:hi_3rod}
\begin{align}
  \bff + \lambda(s) \bfs +  \Lambda(s) \bft  = -\bJf \cdot 
\left[ \frac{\partial \brf}{\partial t}-\buinf 
(\brf)\right],\\
  \bfs + \lambda(s) (\bff + \bft)   = -\bJs 
\cdot 
\left[ \frac{\partial \brs}{\partial t}-\buinf 
(\brs)\right], \\
  \bft + \lambda(s) \bfs +  \Lambda(s) \bff  = -\bJt \cdot 
\left[ \frac{\partial \brt}{\partial t}-\buinf 
(\brt)\right],
\end{align} 
\end{subequations}
where $\Lambda(s) = \ln{\lp H(s)/L \rp}/\ln{\epsilon}$, and $H(s)=2 s 
\sin{\phi}$
denotes the distance between the centreline points $\brf(s)$ and $\brt(s)$  of 
the $1$-st rod
and $3$-rd rod separated by $2\phi$. Abbreviating $\bJ^{(k)} \cdot 
\left[ \partial \br^{(k)}/\partial t - \buinf (\br^{(k)}) \right]$ by 
$\bR^{(k)}$, we obtain the force densities as

\begin{subequations}
\begin{align}
 \bff  & = \frac{(1-\lambda ^2) \bR^{(1)} + \lambda  (\Lambda -1) 
\bR^{(2)} +  \left(\lambda ^2-\Lambda \right)\bR^{(3)}}{(1-\Lambda ) \left(2 
\lambda ^2-\Lambda -1\right)}, \\
 \bfs & =  \frac{(\Lambda +1) \bR^{(2)}-\lambda  (\bR^{(1)}+\bR^{(3)})}{2 
\lambda ^2-\Lambda -1}, \\
\bft & = \frac{ \left(\lambda ^2 - \Lambda\right) \bR^{(1)} + \lambda(  
\Lambda - 1 )\bR^{(2)} +\left(1 - \lambda ^2\right) \bR^{(3)} }{(1-\Lambda) 
\left(2 \lambda ^2-\Lambda -1\right) }.
\end{align}
\end{subequations}

\subsection{Results: Resistance problems}
\begin{figure}
\centering
\includegraphics[scale=0.5]{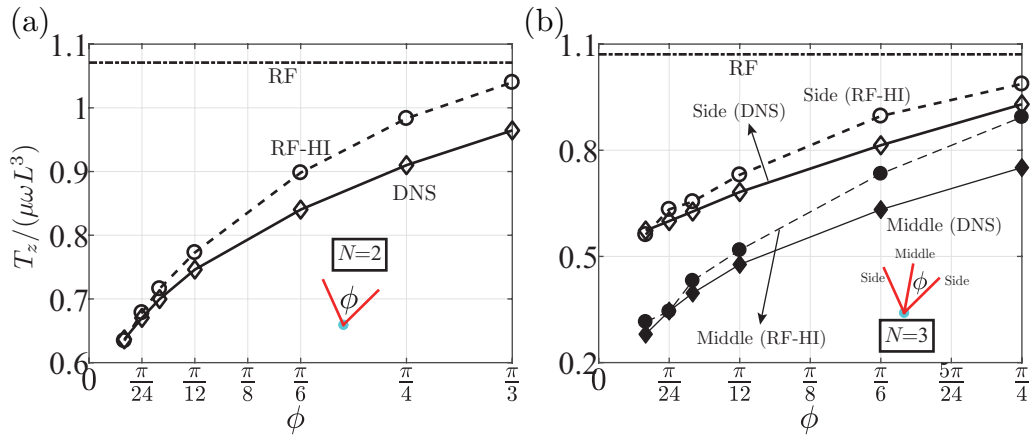}
		\caption{Nondimensional hydrodynamic torque 
$T_z/(\mu \omega L^3)$ exerted on a single rod of a (a) two-rod and (b) 
three-rod watermill 
rotating in the $xy$ plane at a prescribed velocity $\omega$. The torques 
computed by RF, RF-HI theory and DNS  
are shown versus the separation angle 
$\phi$. For three rods, solid/hollow symbols indicate the torques on the 
middle/side rod.}  
\label{fig:2rods_torque_rf_hi}
\end{figure}

We start with the resistance problems. We consider a quiescent environment, 
where 
two or three rods rotate in the $xy$ plane about the joint at a prescribed 
rotational velocity $\omega$. The 
nondimensional hydrodynamic torque $T_z/(\mu \omega L^3 )$ exerted on one rod 
is computed 
by 
RF, RF-HI theory and DNS, and shown in figure~\ref{fig:2rods_torque_rf_hi}. 
For two rods (figure~\ref{fig:2rods_torque_rf_hi}(a)), we observe that when 
they get closer to each other, i.e. $\phi$
decreases, the torque calculated by DNS decreases significantly. This implies
the role of hydrodynamic interactions to reduce the hydrodynamic resistances. 
The RF theory cannot predict any hydrodynamic interactions, while the RF-HI 
predictions agree with the DNS data very well when $\phi \leq \pi/24$. At the 
smallest separation angle, $\phi = \pi/36$, the relative errors of the RF and 
RF-HI predictions are $68\%$ and $0.3\%$, respectively. Increasing $\phi$,
the RF-HI theory deviates from the DNS data, which
is expected considering the $\phi \rightarrow 0 $ asymptotic limit in which 
the RF-HI theory is derived. When $\phi=\pi/3$, the relative errors of the RF 
and RF-HI predictions are $11\%$ and $8\%$; the RF-HI theory fails to
improve the prediction significantly.

In the case of three rods (figure~\ref{fig:2rods_torque_rf_hi}(b)), the 
DNS data shows that the torque on the middle rod 
is less than that on the side rods because the middle rod experiences stronger
hydrodynamic interactions than the other two. The RF-HI theory well 
predicts the hydrodynamic interactions when $\phi \leq \pi/12$. For example, 
for the side 
rod, when $\phi = \pi/36$, the relative 
error  is $87\%$ for the RF prediction, which drops to $2\%$ 
for the RF-HI prediction; when $\phi = \pi/3$, the errors for the two  are 
$15\%$ 
and $6\%$, respectively.

\subsection{Results: Mobility problems}
\begin{figure}
\centering
\includegraphics[scale=0.5]{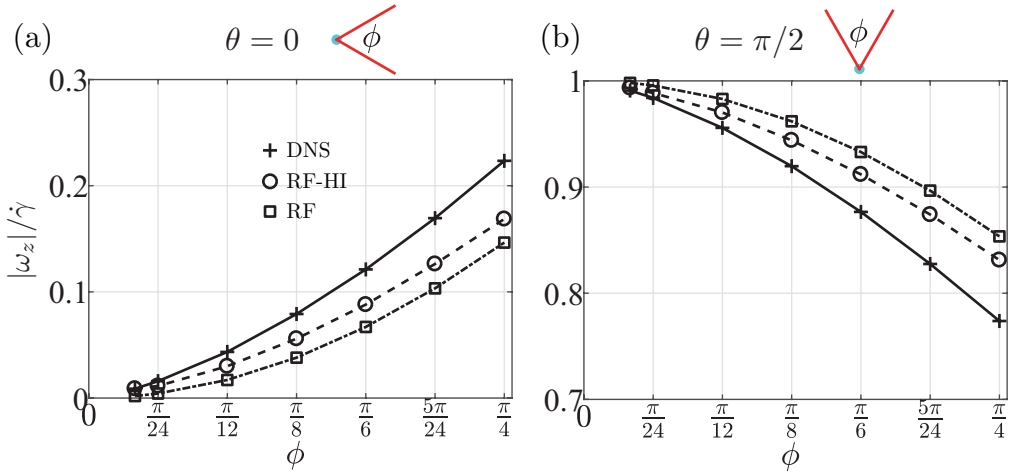}
		\caption{The instantaneous nondimensional 
rotational velocity 
$|\omega_z|/\dot{\gamma}$ of a shear-driven two-rod watermill oriented at (a)
$\theta=0$ and (b) $\theta=\pi/2$, as a function of the separation angle 
$\phi$. The DNS data, RF-HI and RF predictions are compared. }
\label{fig:2rods_omega_shear_rf_hi}
\end{figure}

\begin{figure}
\centering
\includegraphics[scale=0.5]{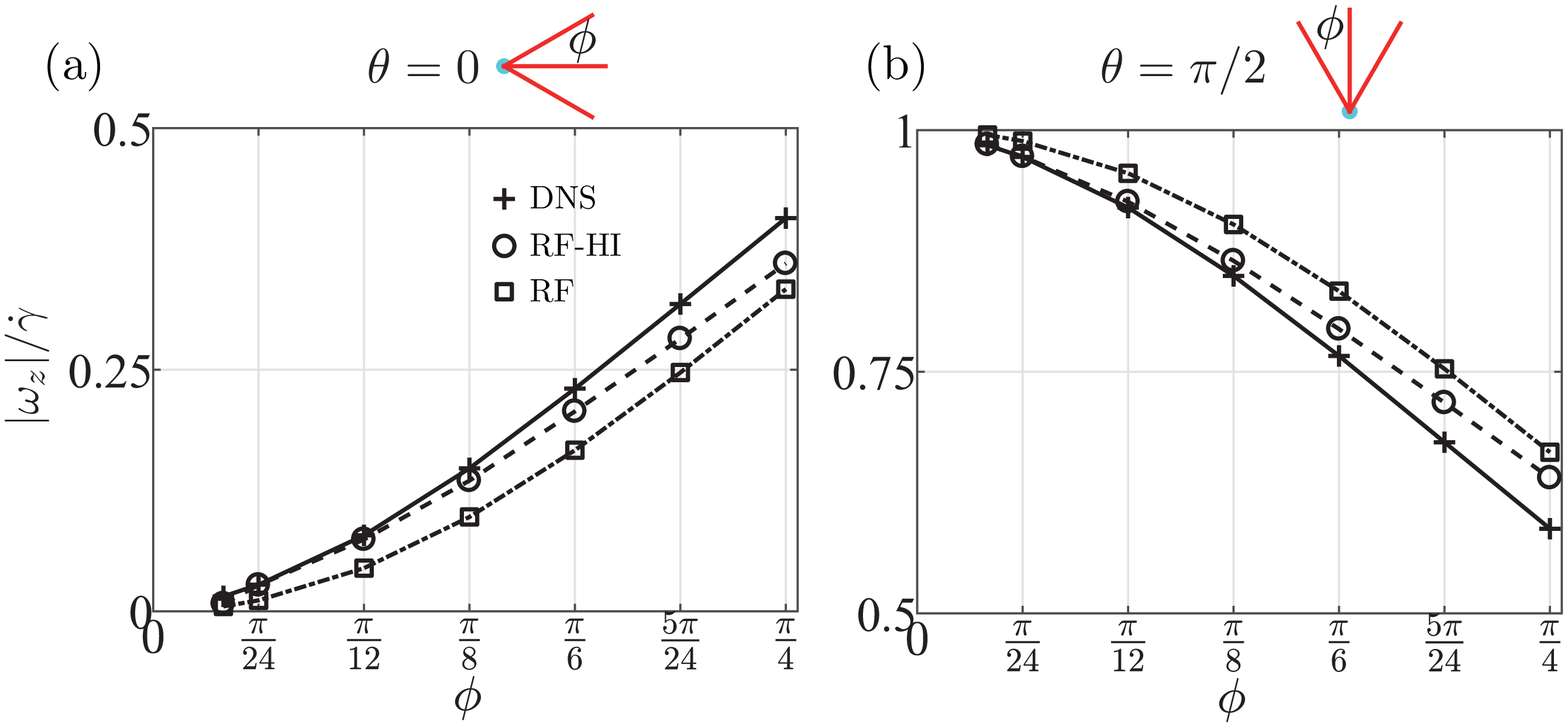}
		\caption{ The instantaneous nondimensional 
rotational velocity 
$|\omega_z|/\dot{\gamma}$ of a shear-driven three-rod watermill oriented at (a)
$\theta=0$  and (b) $\theta=\pi/2$, as a function of the separation angle 
$\phi$. The DNS data, RF-HI and RF predictions are compared. }  
\label{fig:3rods_omega_shear_rf_hi}
\end{figure}

Having demonstrated the resistance problems, we next address the mobility
problems. Namely, for a freely rotating two/three-rod watermill subject to an 
ambient shear flow $\buinf=\dot{\gamma} y \be_x$, where $\dot{\gamma}$ is the 
shear rate, we solve for its rotational 
velocity $\omega_z$ based on the condition that the total hydrodynamic torque
exerted on the watermill is zero. The rotation of the watermill depends on its
orientation, $\theta$, defined here as the angle between the bisector of the 
two rods when $N=2$ or the middle rod when $N=3$ with respect to the $x$-axis 
(see figure~\ref{fig:inter_2rod}(c) and (d)).
For two orientations $\theta=0$ and $\pi/2$, the nondimensional rotational 
velocities $\omega_z /\dot{\gamma}$ versus the separation angle $\phi$
are shown in figure~\ref{fig:2rods_omega_shear_rf_hi} for $N=2$ and 
figure~\ref{fig:3rods_omega_shear_rf_hi} for $N=3$. As expected from 
familiarity
with the Jeffery orbit of a rigid rod subject to a 
shear flow~\citep{jeffery1922motion},
in both cases,
the watermill rotates faster when it is aligned with the shear direction 
($\theta=\pi/2$) than when it is aligned with the flow direction ($\theta=0$).
This feature can be explained by decomposing the shear flow by a clockwise 
rotational 
flow and a planar hyperbolic flow that stretches (resp. compresses)  fluid 
elements 
in the $x=y$ (resp. $x=-y$) direction. Indeed,~\citet{jeffery1922motion} 
determined that a single 
rod under shear attained its maximum rotational velocity $|\omega_z| = 
\dot{\gamma}$ at $\theta=\pi/2$ and minimum $|\omega_z| = 0$ at $\theta=0$. 
Figure~\ref{fig:2rods_omega_shear_rf_hi} shows that in the $\phi \rightarrow 
0$
limit, the two rods merge to one whose rotational velocity indeed approaches 
these two extremes. With increasing $\phi$, the two rods located at 
$\theta=0$ begin to move away from the low rotation region and hence 
$|\omega_z|$ increases with $\phi$; its decrease with $\phi$ when 
they are located
at $\theta = \pi/2$ can be 
explained likewise. The RF theory deviates with the DNS data considerably for 
all of the $\phi$ values, while
the RF-HI agrees with the DNS data when $\phi \leq \pi/24$ but otherwise 
exhibits limited improvement over RF. More specifically, when $\theta=0$, the 
relative errors of the RF and RF-HI predictions for $\phi 
= \pi/36$ are $78\%$ and $0.5\%$, 
respectively; they become $34\%$ and $24\%$ for $\phi = \pi/4$.
The physical picture for $N=3$ presented 
in figure~\ref{fig:3rods_omega_shear_rf_hi} resembles that of $N=2$; a 
notable difference is that the agreement between the RF-HI theory and DNS data
has improved, which does not degrade significantly with $\phi$  until 
$\phi \geq 5\pi/24$.

\begin{figure}
\centering
\includegraphics[scale=0.35]{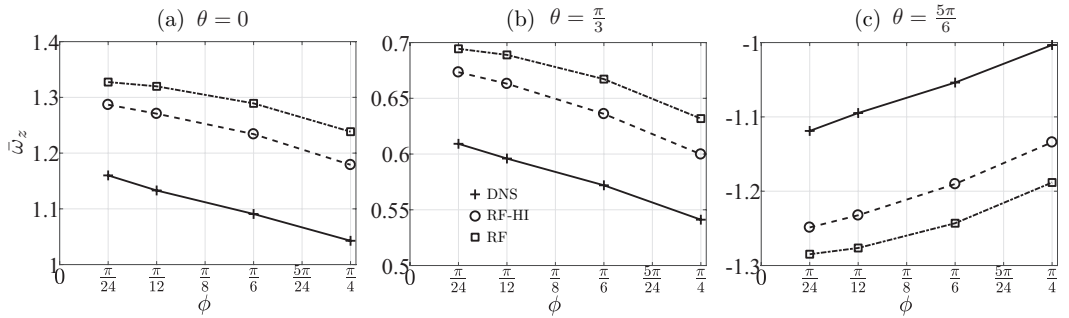}
		\caption{Similar to the setup as in 
figure~\ref{fig:2rods_omega_shear_rf_hi}, but subject to a channel flow 
(instead 
of shear flow) with a characteristic width $\rho=5$ and off-centre 
displacement $\beta=1$, where the watermill is oriented at (a) $\theta=0$, 
(b) $\theta=\pi/3$ and (c) $\theta=5\pi/6$.}  
\label{fig:2rods_omega_channel_rf_hi}
\end{figure}

We finally present in figure~\ref{fig:2rods_omega_channel_rf_hi} the rotational 
velocities $\bar{\omega}_z$ versus $\phi$ of a two-rod watermill oriented at 
(a)
$\theta=0$, (b) $\theta=\pi/3$ and (c) $\theta = 5\pi/6$ in a channel flow 
with a characteristic width $\rho=5$ and off-centre displacement 
$\beta=1$.
For all the orientations, the magnitude of $\bar{\omega}_z$ increases with
the decreasing $\phi$. Both the RF and RF-HI theories predict such a trend
and agree with the DNS data: when $\theta=0$, the 
relative errors of the RF and RF-HI predictions for 
$\phi 
= \pi/36$ are $15\%$ and $11\%$, 
respectively; they are $19\%$ and $13\%$ for $\phi = \pi/4$. Overall, the RF-HI 
theory makes a limited improvement over the RF counterpart.
The results of a three-rod watermill are similar to this two-rod case and hence 
are not reported here.

\section{Conclusions and discussions}\label{sec:conclusion}
Electrically or magnetically driven rotary components are commonly applied 
in microfluidic 
devices for pumping and mixing. In this work, we have performed a theoretical 
and computational study on the hydrodynamics of an anchored
watermill 
as a model passive rotary microfluidic element whose motion is created by the 
flow. This idea was already studied for configurations with a freely rotating 
cylinder in a flow~\citep{day2000lubrication}.
Our work for a 
watermill configuration is motivated 
by the recent experiments of ~\cite{moon2015rotary}, which utilised the 
underlying flow alone to rotate microgears in a microfluidic channel.

We model the watermill as a collection of slender cylindrical rods (like 
paddles) jointed on
a common end, focusing on its hydrodynamic behaviour in an unbounded 
low-Reynolds-number, pressure-driven flow. The classical RF theory linking the 
hydrodynamic 
forces on the rods and their velocities is used to find the relation between 
the hydrodynamic
torque on the watermill and its rotational velocity. By
employing the torque-free condition of the flow-rotated watermill, we obtain
the rotational velocities as a function of the number $N$ of rods, the 
off-centre displacement $\beta$ and orientation $\theta$ of the watermill, and 
the characteristic width 
$\rho$ of the channel flow. For $N \leq 3$, the rotational velocity $\bomgz \lp 
\t \rp = \d{\theta}/\d t$ of the watermill is a function of its orientation 
$\t$. 
By
regarding $\bomgz \lp \t \rp$ as a one-dimensional dynamical system, 
we have analysed its fixed points (indicating the equilibrium orientations of 
the watermill) and their stability. For $N=1$, the single-rod 
watermill will adopt a stable equilibrium orientation that is aligned with the 
downstream flow direction. For $N=2$, one half-stable fixed-point solution 
emerges, corresponding to the configuration that the two-rod watermill is 
parallel with the underlying flow. For $N=3$, when it is placed close enough to 
the centre of the flow, the watermill is aligned with a stable equilibrium 
orientation; otherwise it keeps rotating in flow, implying the absence of fixed 
points when $\beta$ is above a critical value. More interestingly, when $N \ge 
4$, the rotational velocity $\bomgz = 
-\beta/\rho^2$ is independent of  $N$ and $\t$. Namely, the watermill is able 
to rotate at a constant velocity as long as it is not exactly on the 
centreline of the Poiseuille flow. To the best of our knowledge, this striking 
independence has not been reported by the previous studies.  We believe 
that these results will provide a fundamental yet practical guide for the 
future experimental implementation of such flow-driven rotary devices.

It is worth pointing that the independence of the 
rotational velocity predicted by the RF theory reveals the full rotational 
symmetry (that is of infinity order) as long as the geometry of the watermill 
exhibits a rotational symmetry of at least order four, see 
equation~(\ref{eq:omegaz}). We note 
that a referee pointed out this observation exemplifies how a lower-fold 
symmetry may lead to a subtle unexpectedly higher-fold symmetry and it is 
analogous to the full rotational symmetry of the moment of inertia (second 
moment) of a two-dimensional cross section characterised by a
rotational symmetry of at least order three. We comment that the symmetry 
argument of our RF predictions
applies only for this particular configuration and flow, and it holds only in the 
limit of no hydrodynamic interactions. The numerical results indicate that this 
independence is not a general feature of rotationally symmetric objects in 
Stokes flow.

We have developed a  well-validated FEM toolkit based on COMSOL and 
performed DNS to verify the theoretical results. The RF predictions 
are compared with the DNS results and qualitative agreement between them
is observed.
When $N \geq 4$, a systematic underestimation by the RF theory is observed.
This feature seems to imply that the important
role of hydrodynamic interactions neglected by the RF theory needs to be 
considered for a more accurate prediction.
Consequently, an RF-based investigation considering hydrodynamic 
interactions is carried out by leveraging the mathematical framework proposed 
by~\cite{man2016hydrodynamic}; this RF-HI theory is 
asymptotically valid in the small $\phi$ limit (indicating every two 
successive rods are approximately parallel). 

We solve both the resistance and mobility problems of a simpler watermill that 
only consists of two or three closely-spaced rods. For a watermill of two/three 
rods rotating with a 
prescribed velocity in a quiescent flow, the hydrodynamic 
torques predicted  by the RF-HI theory agree  with the numerical results  
quantitatively
when $\phi$ is sufficiently small (within $(0, \pi/24]$), where the relative 
difference is of order $\mathcal{O}(1\%)$. The differences between the 
two increase with $\phi$ as 
expected; nevertheless, hydrodynamic interactions are qualitatively captured 
even for 
$\phi \rightarrow \pi /3$. We then investigate a freely rotating watermill 
driven by shear and Poiseuille flows: for shear flow, the RF-HI theory and DNS 
agree well in the small $\phi$ limit, deviating with each other with 
increasing
$\phi$; for Poiseuille flow, the RF-HI predictions agree better with the DNS 
results than the RF predictions, while no quantitative agreement between 
RF-HI and DNS results is achieved even in the small $\phi$ limit. 

We recognise that the RF-based theories for the hydrodynamic forces/torques on 
slender bodies are not as accurate in a channel flow as compared to the 
quiescent case or shear flow. This fact might be attributed 
to the
strong arclength-dependence of the velocity of the rod relative to the 
underlying flow; in the channel flow, this relative velocity might vary 
its direction along the arclength, which is zero at a particular position
of the rod.  Thus, the force densities in the 
region encompassing the zero-relative-velocity position are 
poorly predicted owing to the failure of RF's basic hypothesis, \textit{viz.}, 
the local force densities linearly depend on the local relative velocities;
in that low-relative-velocity region, the induced 
velocity from the hydrodynamic interaction among different rod segments 
dominates and therefore determines the force 
densities~\citep{johnson1979flagellar}.
The RF-based theories cannot take into account such self interaction and hence
underperform. We infer that the RF-based theories work better when the 
slender structures move in an quiescent environment, uniform flow or a flow
varying slowly in space.

On the computational aspect, it is worth noting that our strategy of using FEM 
to solve mobility problems in the Stokes regime
shares the same convenience  of using the boundary integral method (BIM), 
\textit{viz.}, 
for an instantaneous configuration, the translational/rotational 
velocity of freely translating/rotating objects is obtained by solving 
once a linear system that embeds the force/torque-free condition in the 
discretised form of Stokes equations.
In our experience, this 
FEM approach is more 
computationally expensive than a BIM solver for an unbounded configuration, 
while the overhead reduces significantly for bounded simulations. More 
importantly, this approach naturally offers accurate flow fields that 
are cumbersome to obtain based on BIM solvers.

The flow-driven micro-scale watermill can be applied 
for flow sensing, viz., measuring the local flow rate and/or shear rate based
on the rotational velocity of a watermill. This approach was reported 
in \citet{attia2008modifications}, in analogy with the soft spring 
method~\citep{attia2009soft}. We also expect that such rotary elements 
can be
potentially
used for low-Reynolds-number fluid mixing as a generic and important 
process of microfluidic applications~\citep{whitesides2006origins}. 
It is worth noting that the underlying steady flow becomes unsteady but periodic 
when
perturbed continuously by a rotating watermill. This feature indicates the 
promising potential of this strategy that relies on its capacity to 
introduce 
time-dependent perturbations into the flow ``passively'', which are known to 
generally enhance mixing, instead of actively 
relying on external fields.

\section*{Acknowledgements}
We acknowledge useful discussions with Drs. Yi Man and Sheng Mao, and 
Profs. 
On-Shun Pak, Eric 
Lauga and Fran\c{c}ois Gallaire. We thank the anonymous referees for 
their insightful remarks. L.Z. thanks the Swedish Research Council  for 
a VR International Postdoc Grant (2015-06334). The computer time was
provided by SNIC (Swedish National Infrastructure for Computing).

\appendix

\section{Roots of unity and rotational symmetry}\label{sec:unit_root}
We hereby prove equations~(\ref{eq:unitroot1}), (\ref{eq:unitroot2}) and 
(\ref{eq:unitroot}).
The orientation of the $k$-th ($1 
\leq k \leq N$) rod is
$\theta^{(k)} = \theta + \lp k-1 \rp \phi $ with $\phi = 2\pi /N$. We denote 
the $N$ roots of unity 
satisfying
$w^{N}-1 = 0$ as
\begin{align}
 w_{k-1} = \exp{\left[  i \lp k-1 \rp \phi \right]}, \quad 1 \leq k \leq N.
\end{align}
We write $\cos m \theta^{(k)} = 
\mathfrak{R} \left[\exp{\lp  i m \theta \rp}  w^m_{k-1}\right]$ and
\begin{align}\label{eq:sum_cos}
 \sum_{k=1}^{N}  \cos m \theta^{(k)} = 
\mathfrak{R} \left[\exp{\lp  i m \theta \rp} \sum^{N}_{k=1} w^m_{k-1}\right].
\end{align}
Now we recall the following properties of unit roots
\begin{align}\label{eq:sum_root}
 \sum^{N}_{k=1} w^m_{k-1} = 
 \begin{cases}
  0, & \text{if } m \text{ is not a multiple of } N, \\ 
  N, & \text{if } m \text{ is a multiple of } N,
 \end{cases} 
\end{align}
which physically represents the total $m$-th 
moment of $N$  unit point masses equally spaced on a unit circle; the 
first and second moments represent the centre of mass and moment of 
inertia, respectively. 
By substituting $m=1, 2$ and $3$ into equation~(\ref{eq:sum_cos}) and using 
equation~(\ref{eq:sum_root}), we derive
equations~(\ref{eq:unitroot1}), (\ref{eq:unitroot2}) and (\ref{eq:unitroot}), 
respectively. Note that equation~(\ref{eq:unitroot2}) (second moment) implies 
the 
full rotational symmetry of the moment of inertia of a two-dimensional cross 
section with at least $3$-fold rotational symmetry. On the other hand, 
equation~(\ref{eq:non_tq1}) indicates that the hydrodynamic torque exerted on
an individual rod involves up to the third moment, accordingly, the torque 
presents full rotational symmetry when the watermill is featured by
a rotational symmetry of at least order four.

\end{document}